# Insights into the Mechanism, Selectivity, and Substituent Effects in the Diels-Alder Reaction of Azatrienes with Electron-rich Dienophiles

Amine Rafik,[a,b] Abdeljabbar Jaddi,[a] Mohammed Salah,[c] Najia Komiha,[a] Miguel Carvajal,[b] and Khadija Marakchi*[a]

[a] Laboratory of Spectroscopy, Molecular Modeling, Materials, Nanomaterials, Water and Environment, LS3MN2E/CERNE2D, Faculty of Sciences, Mohammed V University in Rabat, Morocco, k.marakchi@um5r.ac.ma.
[b] Departamento de Ciencias Integradas, Centro de Estudios Avanzados en Física, Matemática y Computación; Unidad Asociada GIFMAN, CSIC-UHU, Universidad de Huelva, Huelva, 21071, Spain.
[b] Molecular Modelling and Spectroscopy Research Team, Faculty of Science, Chouaïb Doukkali University, P.O. Box 20, 24000 El Jadida, Morocco.

The reactivity and mechanistic intricacies of azatrienes in Diels-Alder reactions have been relatively unexplored despite their intriguing potential applications. In this study, we employ Molecular Electron Density Theory to theoretically investigate the hetero-Diels–Alder reaction involving azatrienes with ethyl vinyl ether and allenyl methyl ether. Analysis of Conceptual Density Functional Theory, energetic profiles, and the topological characteristics is conducted to elucidate the reactions. The revealed mechanism manifests as a polar one-step two-stages process under kinetic control. We establish a clear relationship of between the periselectivity, regioselectivity, and stereoselectivity on one hand and the characteristics of the reactions mechanism on the other hand. The influence of weak interactions on reaction activation barriers and bonding evolution are discussed in detail. We demonstrate that substituents enhancing the reverse electron density flux facilitate the feasibility of the reactions. The results lay ground for a meticulous control of the reaction of azatriene in similar synthetic scenarios.

## Introduction

Cycloaddition reactions represent highly efficacious tools for synthesizing a diverse spectrum of carbocyclic and heterocyclic molecules, playing an indispensable role in the contemporary synthesis of biologically active compounds and functional materials, and have industrial and atmospherical applications[1,2]. Paramount in these endeavors is the meticulous control of regiochemistry and stereochemistry. The Hetero-Diels–Alder (HDA) method stands out for its capacity to synthesize a broad range of six-membered heterocyclic compounds. This methodology is particularly promising due to its atom-economic fashion of constructing heterocycles while adeptly managing

chemo-, regio-, and stereoselectivities. Azatrienes emerge as candidates with significant synthetic utility, serving as precursors for generating nitrogen-containing heterocycles. Their possession of multiple reactive sites renders them highly amenable to diverse synthetic modifications. Of particular interest is their application in [4+2] and [2+2] cycloadditions, leading to the formation of corresponding nitrogen heterocycles. Notably, their reactivity in HDA reactions is pronounced when electron-donating groups (-NR2, -OR, -R, ...) are strategically positioned on the nitrogen atom. Conversely, their responsiveness to electron-rich dienophiles is enhanced with the introduction of electron-withdrawing groups (-COR, -COOR, -SO2, ...), rendering them predisposed to engage in inverse electron-demand HDA reactions. Another noteworthy attribute of azatrienes, and more generally of cross-conjugated trienes or dendralenes, is their capability to participate in domino Diels–Alder reactions[3,4]. Domino or tandem reactions denote sequential processes in a series of multistep transformations, and their application has emerged as a highly effective and advantageous strategy in organic synthesis. Indeed, the synthetic efficiency of Diels-Alder (DA) reactions is empowered when they are carried out in a cascade fashion. The diene-transmissive HDA reaction represents one of these methodologies. The theoretical exploration of cycloaddition reactions has undergone remarkable advancements, facilitated by the integration of conceptual Density Functional Theory (CDFT)[5], topological analysis of the electron density[6-10], and Molecular Electron Density Theory (MEDT)[11]. These methodologies have fundamentally transformed our comprehension of reactivity, electronic structure, weak interactions, and bonding dynamics in cycloadditions, and provided a robust foundation for the development of rational design strategies, optimization of reaction conditions, and accurate prediction of reaction outcomes. As these theoretical frameworks undergo ongoing refinement and evolution, the anticipation of further groundbreaking insights in the field is well-founded, propelling innovation in organic synthesis and enriching our chemical toolbox. Despite these substantial advancements, the field of cycloaddition chemistry still knows with persistent and puzzling challenges related to selectivity and mechanistic intricacies[12-15]. Moreover, the quest for a universally consistent methodology to study cycloaddition reactions remains an ongoing challenge, necessitating tailored investigations for distinct classes of cycloadditions or specific organic synthesis scenarios. The mechanistic intricacies of Diels-Alder reactions have sparked extensive debate since the early days of their discovery[16-24]. The fundamental question of whether the changes occurring during the DA process can be accurately described as concerted has long been questioned. Even in the early days of the mechanistic investigations of the DA reaction a two-stage mechanism was proposed by Woodward and Katz[17]. Contemporary computational studies, leveraging precise quantum chemistry methods and tools, have dispelled the widespread acceptance of the synchronous concerted mechanism, commonly found in textbooks, for the majority of possible Diels-Alder reactions. Instead, a prevailing preference emerges for two-stage mechanism and, less frequently, stepwise mechanisms[25-27]. The difference between the one-step and the one-step two-stages mechanisms is not rigorously defined in the literature. Generally, the latter denotes that the formation of the two new bonds occurs within a single kinetic step but in a highly asynchronous fashion, discernible through two distinct stages in the bond formation process[19,28,29]. Domingo et al.[30] propose a definition based on the bond distance of the forming bonds in the transition state (TS), suggesting that bond formation processes in cycloaddition reactions typically occur within the short range of 1.9-2.0 Å. Consequently, a one-step two-stage mechanism is characterized by one new bond distance falling within this range while the other does not. Another perspective introduces a definition based on the time gap between the formation of the two bonds, a more exhaustive but computationally demanding dynamical approach. In this framework, Houk et al.[25] advocate considering a reaction dynamically concerted if the time gap is below 60 fs; otherwise, it is classified as a two-stage or stepwise mechanism. In contrast, Firestone et al.[31] adopt a stricter criterion, deeming a mechanism concerted only when the time gap is precisely 0 fs. The reactivity of azatrienes in HDA reactions is little investigated, in particular, the mechanistic aspects are rarely addressed[32,33]. This study considers the reaction involving the N-sulfonyldivinylmethanimine (Azatriene) with electron-rich dienophiles, namely ethyl vinyl ether (EVE) and allenyl methyl ether (AME).

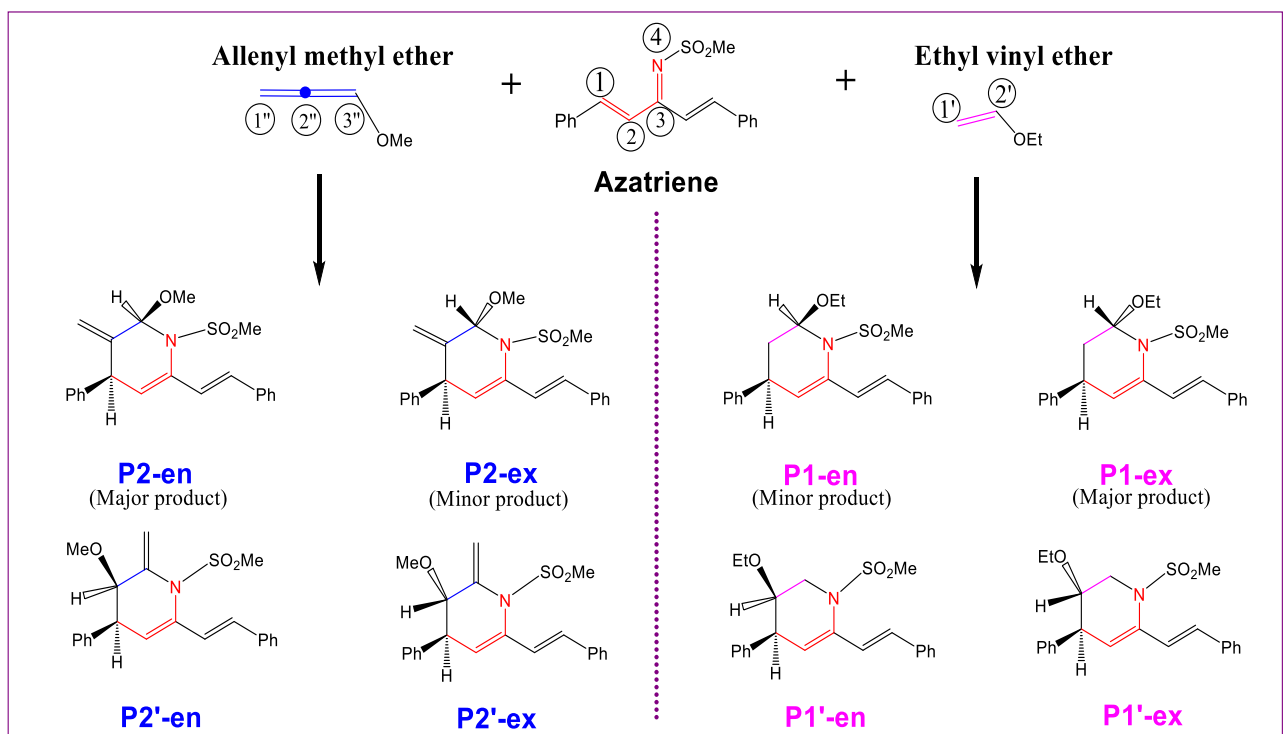

**Figure 1.** General scheme of the reaction of N-sulfonyldivinylmethanimine with ethyl vinyl ether and allenyl methyl ether

Experimental investigations on these reactions have been reported by Kobayashi et al.[34,35]. The Azatriene under consideration is substituted with an electron-withdrawing group, SO2Me, which has the potential to engage in weak interactions with approaching dienophiles owing to its electronic and structural properties. The theoretical exploration of the Azatriene reaction with EVE suggests four potential reaction pathways, while experimental observations reveal a complete regioselectivity, with the major product favoring the endo approach. In contrast, the reaction of Azatriene with AME presents eight potential reaction pathways, stemming from two main reaction routes associated with the periselectivity of the reaction. Experimental findings indicate complete periselectivity, with the major product favoring the exo cycloadduct. Figure 1 illustrates a general scheme of the reaction, including atom numbering for relevant atoms and the experimental major and minor products. Our objective is to comprehend the observed reactivity trends, rationalize the peri-, regio-, and stereoselectivities implicated, and elucidate the underlying reaction mechanisms. The outcomes of this investigation contribute to a deeper understanding of azatriene reactivity in the HDA reactions and aid in predicting their behavior in analogous synthetic scenarios. Additionally, we assess the performance of the computational tools employed in this study.

## Computational and theoretical methods

The reaction pathways were explored using investigated using Density Functional Theory (DFT) at the B3LYP/6-311G(d,p) level of theory, employing the Berny analytical gradient optimization method[36] as implemented in Gaussian09[37]. All subsequent calculations were carried out at the same level of theory. The authenticity of all stationary points was confirmed through frequency calculations, where harmonic frequency analysis revealed no imaginary frequencies for minima and one imaginary frequency corresponding to the bonding vibration in the transition states. Additionally, transition vectors analysis was performed to ensure that the components undergoing the most significant variation are associated with the formation of the new σ bonds. To further validate each reaction pathway, Intrinsic Reaction Coordinate (IRC) calculations[38] were conducted to confirm that the TS is well-connected to the minima of reactants and products along the anticipated pathway. All calculations were performed both in the gas phase and in toluene solvent. In the latter case, the simulation of experimental conditions involved the use of a self-coherent reaction field (SCRF)[39] based on a polarizable continuum model (PCM)[40] at a temperature of 383.15 K. Natural Bond Orbital (NBO) analysis was also carried out with Gaussian09[41]. Conceptual Density Functional Theory (CDFT)[5,42] was employed to extract chemically relevant information from the electronic density and quantitatively describe key chemical concepts, including chemical potential (μ)[43], hardness (η)[43], nucleophilicity (N)[44], and electrophilicity (ω)[42]. These indices serve as powerful tools for examining electronic density flow and the polarity of the reaction. In this framework, the aforementioned chemical concepts are defined as μ = ($E_{HOMO}$+$E_{LUMO}$)/2, η = $E_{LUMO}$ - $E_{HOMO}$, ω = $μ^2$/2η, and N = $E_{HOMO}$ - $E_{HOMO}$ (TCE) respectively. The $E_{HOMO}$ (TCE) is the energy of the HOMO of tetracyanoethylene calculated to be -9,37 eV at the employed level of theory, TCE is conventionally used as a reference due to its known very low HOMO energy. Parr functions[45] were computed to perform analysis of the atomic spin density and reveal the nucleophilic and electrophilic atomic centers in the reactants. These local indices are crucial for predicting the regioselectivity of the reactions. The parr functions were computed using an analysis of the Mulliken atomic spin density of the radical anion and the radical cation of the neutral molecule using the following equations: $P_k^- = \rho_s^{rc}(k)$, $P_k^+ = \rho_s^{ra}(k)$. From that, the local nucleophilicity index $N_k$ and local electrophilicity index $\omega_k$ were expressed as $N_k = NP_k^-$, $\omega_k = \omega P_k^+$. In order to further evaluate the electron demand and polarity of the reaction, the global electron density transfer (GEDT)[30,46] was calculated at the TSs as the sum of natural atomic charges of each fragment of the TS, with the natural atomic charges having been extracted from the Natural population analysis (NPA). The computation of NPA charges solely consists of evaluating the natural atomic orbitals (NAOs) and then summing across all NAOs belonging to a specific atom to derive its Natural Charge. A topological analysis of the electron density is performed for the purpose of characterizing the inter-fragment interactions and primitive changes throughout reaction pathways. Namely, we employ the Atoms-in-Molecules (AIM) theory[47], the Independent Gradient Model based on Hirshfeld partition of molecular density (IGMH)[48], and the Electron Localization Function (ELF) method[7,49]. The AIM theory allows for the division of electron density in the molecular space into regions linked to specific atoms, known as basins. It permits the identification of important interatomic interactions by analysis of the magnitude of the electron density and the sign of its Laplacian. The IGMH is the latest variant of the IGM approach[10] which provides a molecular signature of the various interactions between two fragments of a molecule. In particular, the IGMH variant is shown to be efficient in detecting weak interactions in virtue of the definition of the actual electron density within its formalism. In the IGM approach, a physical descriptor expressed by δg is defined to measure the electron sharing in regions of space where atomic electron densities interfere. This descriptor is computed in the framework of IGMH according to the expression $\delta g = \left|\nabla_\rho^{IGM}\right| - \left|\nabla_\rho\right|$ where $\left|\nabla_\rho\right|$ is the gradient of the electron density, $\left|\nabla_\rho^{IGM}\right|$ is the virtual upper limit of the electron density gradient, and the atomic densities are obtained using a Hirshfeld partition scheme. The IGM method inherently enables the definition of a bond strength score that probes the strength of interaction of a given atom pair by isolating and integrating their signature from the broad interaction signal[50]. We make use of the Intrinsic Bond Strength Index for Weak interactions (IBSIW) implemented in Multiwfn package[51] to quantitively compare the interaction strengths of the forming bonds at the TSs. For a pair of atoms i and j the IBSIW is obtained using the equation:

$$IBSIW(i,j) = 100 \times \frac{\delta G_{i,j}^{pair}}{d_{i,j}^2}$$

Where $\delta G_{i,j}^{pair}$ quantifies the contribution of the atomic pair to the interaction between two fragments. The ELF approach follows a different approach to the topology of the electron density where this latter is localized and integrated[7]. Partitioning of the molecular space using the gradient field of the ELF produces basins where analysis of their population (obtained by integrating the density in their region) and volumes permits visualizing the electron distribution in a molecule using the Lewis's representation of bonding pairs and lone pairs. The ELF function is given by:

$$\text{ELF} = \left[1 + \left(\frac{\tau_\sigma - \frac{(\nabla\rho_\sigma)^2}{4\rho_\sigma}}{\frac{3}{5}(6\pi^2)^{2/3}(\rho_\sigma^{5/3})}\right)^2\right]^{-1}$$

The IGMH was realized using version 3.8 of the Multiwfn package[51], and ELF was realized using the Topmod package[52]. Throughout this work, visualization of the results was carried out using GaussView 6.0[53] and Visual Molecular dynamics (VMD) package[54].

## Results and discussion

This study is organized as follows: The initial subsection involves a detailed examination, featuring a CDFT analysis and an ELF topological analysis of the reactants. This analysis is conducted to discern the nature of electronic flux, polarity, regioselectivity, and periselectivity inherent in the reaction. The subsequent subsection is dedicated to an in-depth exploration of potential reaction pathways for each of the reactions of the **Azatriene** with **EVE** and **AME**. Within this subsection, a comprehensive discussion unfolds regarding the energetics of the reaction, its consequential impact on regio- and stereoselectivities, and the reactions mechanism. In the third subsection, we elucidate comprehensively the underlying mechanisms of the studied reactions, employing an investigation into charge transfer phenomena, Wiberg bond orders, and the topology of the electron density within the TSs.

### 1. Interactions between the reactants

### 1.1. Global CDFT indices

This investigation incorporated the calculation of global CDFT indices, which are detailed in Table 1. The chemical potentials of **EVE** (-2.75 eV) and **AME** (-3.23 eV) surpass that of **Azatriene** (-4.69 eV). This disparity implies that the reactions possess an inverse electron demand (IED) nature per the classical classification of these reactions[55,56]. Later, in the GEDT section we will categorize the reaction based on the more chemically meaningful concept of forward and reverse electron density flux[57]. Moreover, Azatriene exhibits pronounced global electrophilicity (2.98 eV) on the scale proposed by Domingo et al.[58], exceeding that of **EVE** (0.55 eV) categorized as a marginal electrophile, and **AME** (0.88 eV) classified as a moderate electrophile. Consequently, in these reactions, the diene functions as the electrophile, while the dienophiles act as robust nucleophiles, in accordance with the scale established by Jaramillo et al.[59]. Notably, the global nucleophilicities of **EVE** and **AME** surpass 3 eV. Collectively, these findings reinforce the result that the electronic flux takes place from the dienophile to the diene. Nucleophilicity and electrophilicity may be treated as two extreme ends of the one scale. The comparison of the nucleophilic/electrophilic characteristic of a reactant can be performed by assessing either its electrophilicity index, its nucleophilicity index, or more commonly, accounting for both characteristics simultaneously. Typically, reactions between strong electrophiles and strong nucleophiles involve intense density transfer, therefore, they are considered easier. The noticeable electrophilic character of Azatriene and the nucleophilic character of the dienophiles, coupled with the substantial difference in chemical potential between the diene and the dienophiles (1.94 and 1.46 for **EVE** and **AME**, respectively), implies a polar nature of the reactions. However, it is noteworthy that Azatriene concurrently displays a nucleophilic behavior, evident in its relatively high nucleophilicity index of 2.88 eV, which suggests a nuanced modulation of reaction polarity.

**Table 1.** Global reactivity indices for the reaction of Azatriene with EVE and AME. Azatriene-R where R=PhSO$_2$, p-TolSO$_2$, PhCO, Ph, NME$_2$, Me, i-Pr signifies the nitrogen atom of the diene is substituted by the R group.

|  | HOMO (eV) | LUMO (eV) | μ | η | ω | N |
|---|---|---|---|---|---|---|
| **EVE** | -6.16 | 0.66 | -2.75 | 6.82 | 0.55 | 3.25 |
| **AME** | -6.21 | -0.26 | -3.23 | 5.95 | 0.88 | 3.16 |
| *Azatriene* | -6.53 | -2.84 | -4.69 | 3.69 | 2.98 | 2.88 |
| *Azatriene-PhSO$_2$* | -6.50 | -2.80 | -4.65 | 3.69 | 2.93 | 2.91 |
| *Azatriene-p-TolSO$_2$* | -6.47 | -2.77 | -4.62 | 3.70 | 2.88 | 2.94 |
| *Azatriene-PhCO* | -6.37 | -2.55 | -4.46 | 3.82 | 2.60 | 3.04 |
| *Azatriene-Ph* | -5.68 | -2.35 | -4.01 | 3.33 | 2.41 | 3.73 |
| *Azatriene-NMe$_2$* | -5.26 | -2.11 | -3.68 | 3.15 | 2.15 | 4.15 |
| *Azatriene-Me* | -5.96 | -2.00 | -3.98 | 3.96 | 2.00 | 3.45 |
| *Azatriene-i-Pr* | -5.94 | 0.30 | -2.82 | 6.24 | 0.64 | 3.47 |

Additionally, we substituted the MeSO$_2$ group on the nitrogen atom of the diene with a range of substituents (PhSO$_2$, p-TolSO$_2$, PhCO, Ph, NMe$_2$, Me, and i-Pr) to investigate the impact on the electrophilicity and nucleophilicity indices in the dienes, along with its consequential effect on the feasibility of the reaction. The evolution of the electrophilicity and nucleophilicity indices is visually presented in **Figure 2**, while the set of corresponding global indices is reported in **Table 1**.

The electrophilicity scale shows that all examined dienes are strong electrophiles, the only exception corresponds to the i-Pr substituent. **EVE** on the other hand is a marginal electrophile. Therefore, from an electrophilicity standpoint, we anticipate the reactions to proceed smoothly for all N-substituted Azatrienes except for Azatriene-i-Pr, where the reactions are expected to be progressively more challenging due to its low electrophilicity. However, since electrophilicity and nucleophilicity are not united in a single scale, a thorough examination of the nucleophilicity of the reactants is imperative for a comprehensive understanding. In this context, we observe that **EVE** can function as a potent nucleophile, while Azatriene, Azatriene-PhSO$_2$, and Azatriene-p-TolSO$_2$ exhibit moderate nucleophilic behavior. Azatriene-PhCO borders between a moderate and strong nucleophile, whereas the remaining dienes emerge as strong nucleophiles. Considering this spectrum, we anticipate that **EVE** will to react robustly as a strong nucleophile with Azatriene, Azatriene-PhSO$_2$, and Azatriene-p-TolSO$_2$, will exhibit a more challenging reaction with Azatriene-PhCO due to this latter's strong nucleophilic character. The rest of the dienes, particularly Azatriene-i-Pr, characterized by strong nucleophilicity and comparatively weaker electrophilicity, are expected to show minimal to no reactivity at all. These predictions align seamlessly with experimental observations, particularly noting that the reaction of **EVE** with Azatriene-PhCO was successful only in the presence of a catalyst. Overall **AME** displays a similar nucleophilic and electrophilic behavior to **EVE**. We anticipate its readiness to react with MeSO$_2$, Azatriene-PhSO$_2$, and Azatriene-p-TolSO$_2$. Given its specific electrophilicity and nucleophilicity values (0.88 eV and 3.16 eV, respectively), there is a possibility that it reacts with Azatriene-PhCO under harsh conditions, while reactions with the remaining dienes are deemed unlikely. Experimental results validate our predictions for the MeSO$_2$, Azatriene-PhSO$_2$, and Azatriene-p-TolSO$_2$ substituent cases, although the remaining scenarios were not explored by the authors[34,35].

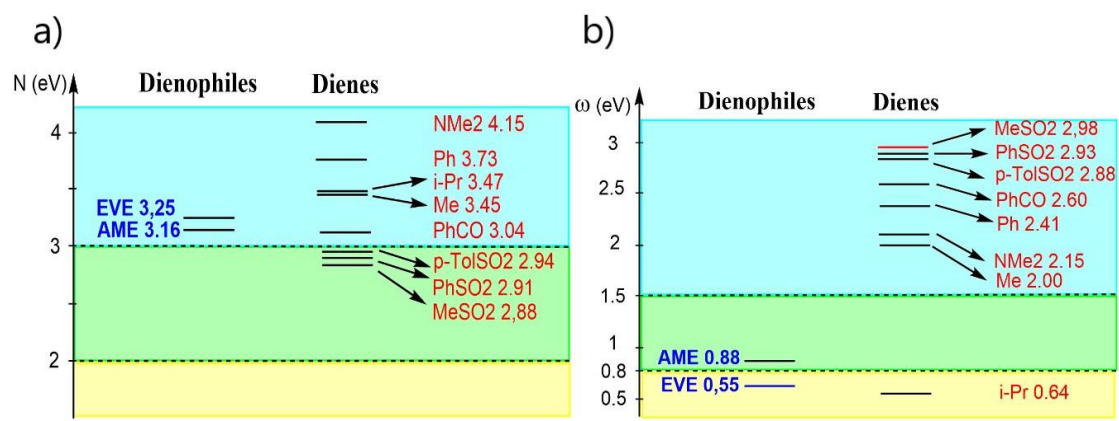

**Figure 1.** Nucleophilicity (a) and electrophilicity (b) classification of the dienophiles and the diene Azatriene and various N-substituted Azatrienes. Yellow, green, and blue stand for weak, moderate and strong electrophilicity/nucleophilicity regions respectively.

## 1.2. Local CDFT indices

The prediction of the atomic sites with the highest likelihood of bond formation from each of the two reactants in a reaction is facilitated through the analysis of Atomic Spin Density via the Parr functions. Local Conceptual Density Functional Theory (CDFT) indices corresponding to these predictions are detailed in **Table 2**. The numbering of the atomic sites is given in Figure 1.

In the reaction of **Azatriene** with **EVE**, the most electrophilic center of Azatriene is identified as C1 ($\omega_k$=0.68), while the most nucleophilic center of the dienophile is C1' ($\omega_k$=1.86). Conversely, the most nucleophilic center of **Azatriene** is N4 ($N_k$=1.05), with the most electrophilic center of the dienophile being C2' ($N_k$=0.17). As a result, the favorable bonding pattern involves the formation of the C1-C1' and N4-C2' bonds, corresponding to the **P1-en** and **P1-ex** in Figure 1. For the reaction of **Azatriene** with **AME**, the reaction may occur in either of the two adjacent double bonds of the **AME** molecule. Considering the three carbon atoms forming these double bonds, the most electrophilic center of Azatriene, C1, bonds with the most nucleophilic center of **AME**, C2'' ($N_k$=1.59). The second most electrophilic center, N4, bonds with the corresponding second most nucleophilic center of **AME**, which is C3'' ($N_k$=0.36). Consequently, the reaction preferentially occurs in the C2''-C3'' bond rather than the C2''-C1'' bond, indicating periselectivity to the C2''-C3''. Furthermore, it is anticipated that the major regioisomer corresponds to the bonding of N4-C2'' and C1-C3'', aligning with experimental results. Regarding the regioselectivity for other N-substituted Azatrienes, the reactions of **EVE** with azatrienes substituted with PhSO$_2$, p-TolSO$_2$, and PhCO follow a similar trend to that of **EVE** and the reference Azatriene. However, For the rest of dienes, we have already predicted that the reactions are not feasible. This is manifested in one way in the N4 atomic center being simultaneously the most electrophilic center and the most nucleophilic center.

**Table 2.** Local reactivity indices for the reaction of Azatriene and N-substituted Azatriene with **EVE** and **AME**.

|     | Atomic Site | P+k   | P-k   | ωk    | Nk    |
| --- | ----------- | ----- | ----- | ----- | ----- |
| **EVE** | C1'     | 0,24  | 0,57  | 0.13  | 1.86  |
|     | C2'         | 0,27  | 0,06  | 0.15  | 0.19  |
| **AME** | C1''    | 0.51  | -0.07 | 0.45  | -0.22 |
|     | C2''        | 0.29  | 0.50  | 0.26  | 1.59  |
|     | C3''        | -0.02 | 0.11  | -0.02 | 0.36  |
| **Azatriene** | C1 | 0.23 | -0.06 | 0.68 | -0.16 |
| | N4 | 0.17 | 0.37 | 0.52 | 1.05 |
| **Azatriene-PhSO2** | C1 | 0.23 | -0.05 | 0.67 | -0.14 |
| | N4 | 0.17 | 0.36 | 0.48 | 1.06 |
| **Azatriene-p-TolSO2** | C1 | 0.23 | -0.04 | 0.66 | -0.13 |
| | N4 | 0.17 | 0.34 | 0.48 | 1.01 |
| **Azatriene-PhCO** | C1 | 0.20 | -0.04 | 0.51 | -0.11 |
| | N4 | 0.18 | 0.37 | 0.48 | 1.11 |
| **Azatriene-Ph** | C1 | 0.17 | 0.01 | 0.41 | 0.01 |
| | N4 | 0.24 | 0.36 | 0.57 | 1.34 |
| **Azatriene-Me** | C1 | 0.11 | 0.01 | 0.23 | 0.04 |
| | N4 | 0.27 | 0.36 | 0.53 | 1.23 |
| **Azatriene-NMe2** | C1 | 0.15 | 0.01 | 0.32 | 0.04 |
| | N4 | 0.31 | 0. 29 | 0.66 | 1.19 |
| **Azatriene-i-Pr** | C1 | 0.09 | 0.01 | 0.06 | 0.05 |
| | N4 | 0.28 | 0.411 | 0.18 | 1.42 |

### *1.3. Topological analysis at the ground state of reagents*

The electronic structure of the reactants is characterized by carrying out a topological analysis via the ELF method. The results, namely, ELF localization domains, basin attractor positions, and the most relevant valence basin populations are visualized in **Figure 3**. On the one hand, Inspection of the ELF basin attractors in the reactive region of **Azatriene** shows one monosynaptic valence basin on the Nitrogen atom V(N) with a large population of 3.02 due to the electronegativity of the Nitrogen atom, and three disynaptic basins for each of the three bonds (C1,C2), (C2,C3), and (C3,N4) with populations 3.27 e, 2.34 e, and 2.72 e respectively. It is noteworthy that the (C3,N4) has a higher population than expected due to the participation in the conjugated π system, i.e., the donor effect of N. And for the very same reason, the (C1,C2) bond region exhibits only one disynaptic basin of large population (3.27 e) rather than two equivalent and much less populated disynaptic basins as we would expect for a double bond. On the other hand, the adjacent π system shows two equivalent disynaptic basins V(C5,C6) of populations 1.62 and 1.67 (in total 3.29 e), this behavior may be attributed to the competition between the two conjugated systems which both share the middle C=N double bond. It is easy to relate these findings to the concept of cross-conjugation where each of the terminal π bonds in a cross-conjugated molecule is conjugated with the central π bond but not with other terminal π bond. With regard to **EVE**, the double bond region is characterized with two matching disynaptic bonds V(C1',C2') of populations integrating 1.81 e and 1.79 e respectively. As for the **AME**, the double bonds candidate to react with **Azatriene** each exhibit two equivalent disynaptic basins. Two V(C1'',C2'') basins integrating populations of 1.81 e and 1.82 e, and two V(C2'',C3'') basins each integrating 1.98 e. We notice that the bond C2''=C3'' (in total 3.96 e) is more populated than the C1''=C2'' (3.63 e). This can be attributed to the donor effect of the oxygen atom, and as a result, the C2''=C3'' is more prone to undergo the HDA reaction with Azatriene than the C1''=C2'' bond, in harmony with the Parr functions analysis results. Given these results, for the remainder of this work, we consider only the reaction of **Azatriene** with **AME** at the C2''=C3'' bond.

The NPA charges are calculated and given in the third column of Figure 3. The atoms C1, C2, C3, and N4 of the **Azatriene** have charges -0.10, -0.26, +0.30, and -0.77 respectively. **EVE** exhibits charges of -0.47 and +0.16 on its C1' and C2' carbons. **AME** exhibits charges of -0.40, -0.07, and +0.09 on its C1'', C2'', and C3'' carbons.

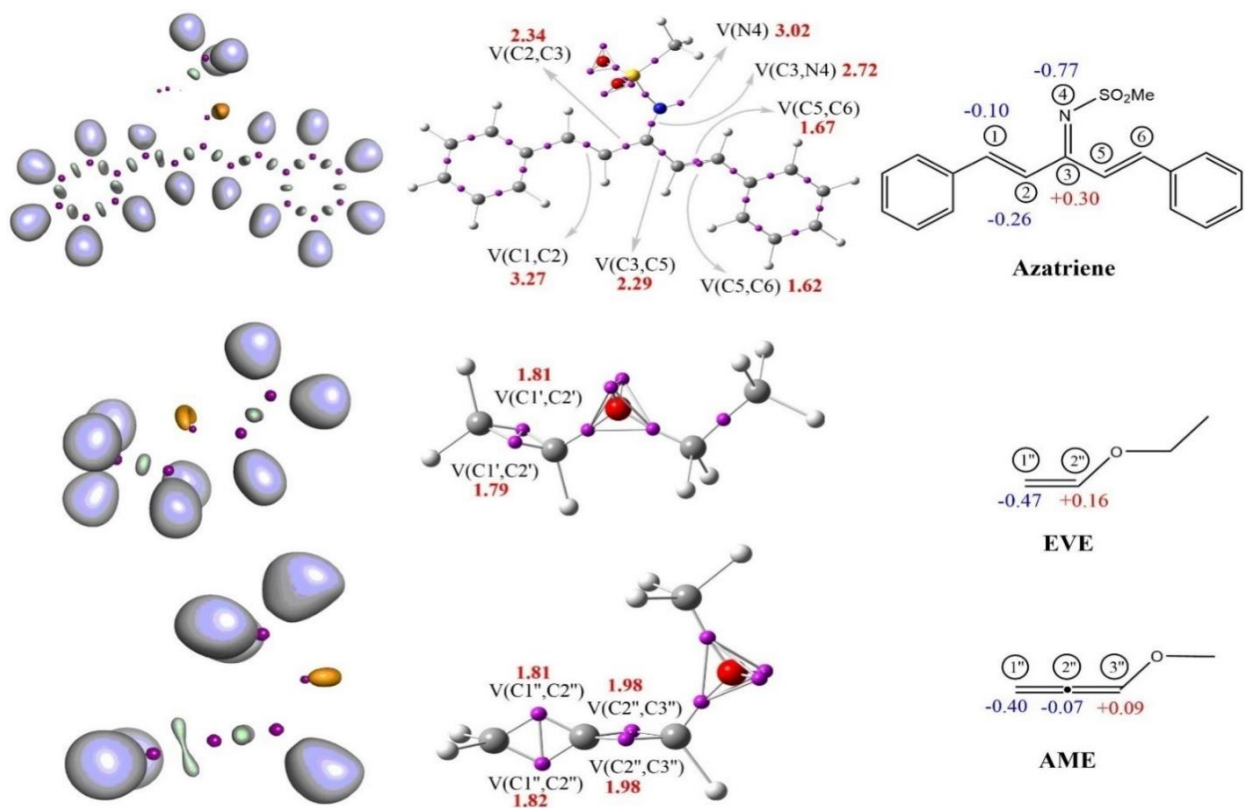

**Figure 2.** (Left column) ELF localization domains of Azatriene and **EVE** (displayed from the upper row to the bottom one, respectively) plotted at an isosurface value of 0.8. Protonated basins are shown in light purple, monosynaptic basins in orange, disynaptic basins in green and core basins in purple. (Middle column) ELF basin attractor positions along with population values for the most relevant sites. (Right column) Lewis-like structures of reactants together with natural atomic charge on relevant atoms.

## 2. Reaction energy profiles
### 2.1. Reaction of Azatriene with EVE

The reaction of **Azatriene** with **EVE** may proceed according to four possible pathways relative to two regioisomers each presenting an endo and an exo stereoisomer (**Figure 1**). The energies of the of the corresponding stationary points in gas phase and in toluene as solvent are summarized in **Table 3**, and the energy profiles are presented in **Figure 4**.

In this Figure, it can be highlighted that the product **P1-en** is kinetically favored with the lowest energy barrier for **TS1-en** at +22.99 kcal/mol, closely followed by the **P1-ex** product with an energy barrier of +23.15 kcal/mol. The transition structures **TS1'-en** and **TS1'-ex** corresponding to the second regioisomer are located at higher energy levels, the energetic difference between them and **TS1-en** are 7.26 kcal/mol and 8.52 kcal/mol respectively. This indicates that, from the kinetic perspective, the reaction is regioselective towards the first regioisomer associated with **TS1-en** and **TS1-ex**. Thermodynamically, the second regioisomer is favored as shown by the energies of **P1'-en** and **P1'-ex** at -19.67 kcal/mol and -19.65 kcal/mol respectively. Nevertheless, **P1-en** is also thermodynamically accessible because the relative energy difference is of only 2.22 kcal/mol compared to **P1'-en**. We except that under kinetic conditions a mixture of **P1-en** and **P1-ex** is to be obtained with **P1-en** being the major product. Stereoselectivity wise, the reaction is stereoselective towards the endo product. The patterns observed in the gas phase are conserved in the toluene solvent, although some small shifts in energy are observed. The regio- and stereoselectivity conclusions are in accordance with the experimental observations.

Table 3. Total energies (a.u) and relative energies (kcal/mol) of reactants, transition states, and products of reaction of **Azatriene** with **EVE** in gas phase and in toluene solvent.

|  | Gas phase | | Toluene | |
|---|---|---|---|---|
|  | E (u.a.) | ΔE (Kcal/mol) | E (u.a.) | ΔE (Kcal/mol) |
| **Azatriene** | -1299.6955973 | - | -1299.7038693 | - |
| **EVE** | -232.4987220 | - | -232.5004698 | - |
| **P1-en** | -1532.222127 | -17.45 | -1532.230250 | -16,26 |
| **P1-ex** | -1532.215610 | -13.36 | -1532.222890 | -11,64 |
| **P1'-en** | -1532.225661 | -19.67 | -1532.232997 | -17,98 |
| **P1'-ex** | -1532.225635 | -19.65 | -1532.233054 | -18,02 |
| **TS1-en** | -1532.157684 | 22.99 | -1532.168747 | 22,33 |
| **TS1-ex** | -1532.157431 | 23.15 | -1532.167273 | 23,30 |
| **TS1'-en** | -1532.146107 | 30.25 | -1532.155226 | 30,80 |
| **TS1'-ex** | -1532.144100 | 31.51 | -1532.153237 | 32,10 |

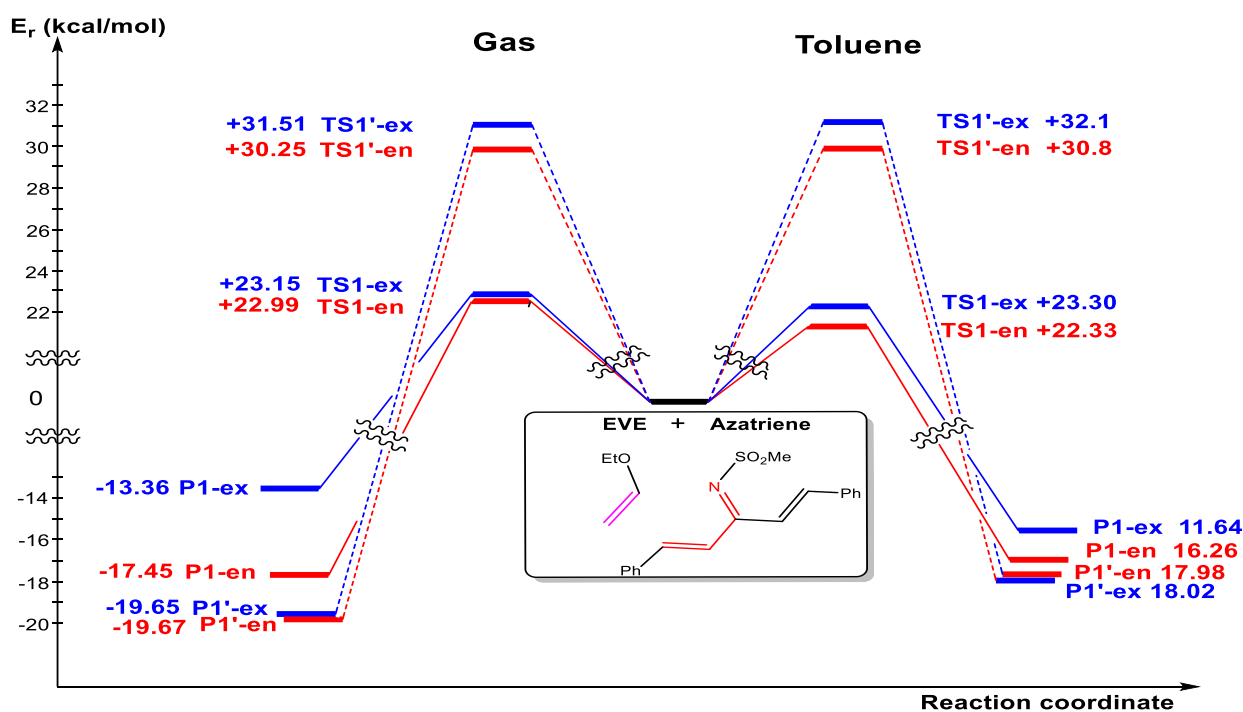

Figure 3. Activation energy profiles for the HDA reaction of **Azatriene** with **EVE** in gas phase and in toluene.

## 2.2. Reaction of Azatriene with AME

Following the results of the CDFT indices and topological analysis, we consider the reaction of Azatriene at the C2''=C3'' bond. Similar to the previous reaction with **EVE**, a total of four reaction pathways can be expected. These pathways correspond to two regio isomers and two stereoisomers arising from the orientation of approach of the addends (see **Figure 1**). The corresponding energies in gas phase and in solvent are shown in **Table 4** and **Figure 5**. On the kinetic level, it can be noted that the reaction is regioselective. Indeed, **TS2-ex** is more stable by approximately 11 kcal/mol compared to the regioisomeric set of **TS2'-en** and **TS2'-ex**. Moreover, the exo product corresponding to **TS2-ex** has a lower activation barrier than the corresponding endo TS, implying that the reaction is kinetically exo stereoselective. Therefore **TS2-ex** is the favored kinetic product. From a thermodynamics perspective, the stereoselectivity is reversed, for this case, it favors the endo product. **P2-en** is the most stable product, and **P2-ex** is less stable by 1.87 kcal/mol. Consequently, **P2-en** is the thermodynamic product. No significant changes are observed when the solvent and temperature effects are considered, the relative order of stationary points remains the same. Overall, under kinetic conditions, the results regarding regioselectivity and stereoselectivity are in agreement with the experimental observations. should be noted that through all reactions involving **Azatriene** investigated in this study, the comparison with the experiment indicates that the reactions are under kinetic control.

**Table 4.** Total energies (a.u) and relative energies (Kcal/mol) of reactants, transition states, and products of the reaction of **Azatriene** with **AME** in gas phase and in toluene solvent.

|  | Gas phase |  | Toluene |  |
| --- | --- | --- | --- | --- |
|  | E (a.u) | ΔE (Kcal/mol) | E (u.a.) | ΔE (Kcal/mol) |
| **Azatriene** | -1299.6955973 | - | -1299.7038693 | - |
| **AME** | -231.24193990 | - | -231.24341770 | - |
| **P2-en** | -1530.9828869 | -28.46 | -1530.9906930 | -27.24 |
| **P2-ex** | -1530.9799129 | -26.59 | -1530.9882263 | -25.69 |
| **P2'-en** | -1530.9713822 | -21.24 | -1530.9796083 | -20.28 |
| **P2'-ex** | -1530.9760069 | -24.14 | -1530.9838630 | -22.95 |
| **TS2-en** | -1530.8982483 | 24.65 | -1530.9089460 | 24.06 |
| **TS2-ex** | -1530.9037490 | 21.20 | -1530.9137095 | 21.07 |
| **TS2'-en** | -1530.8855589 | 32.61 | -1530.8949451 | 32.84 |
| **TS2'-ex** | -1530.8862608 | 32.17 | -1530.8944422 | 33.16 |

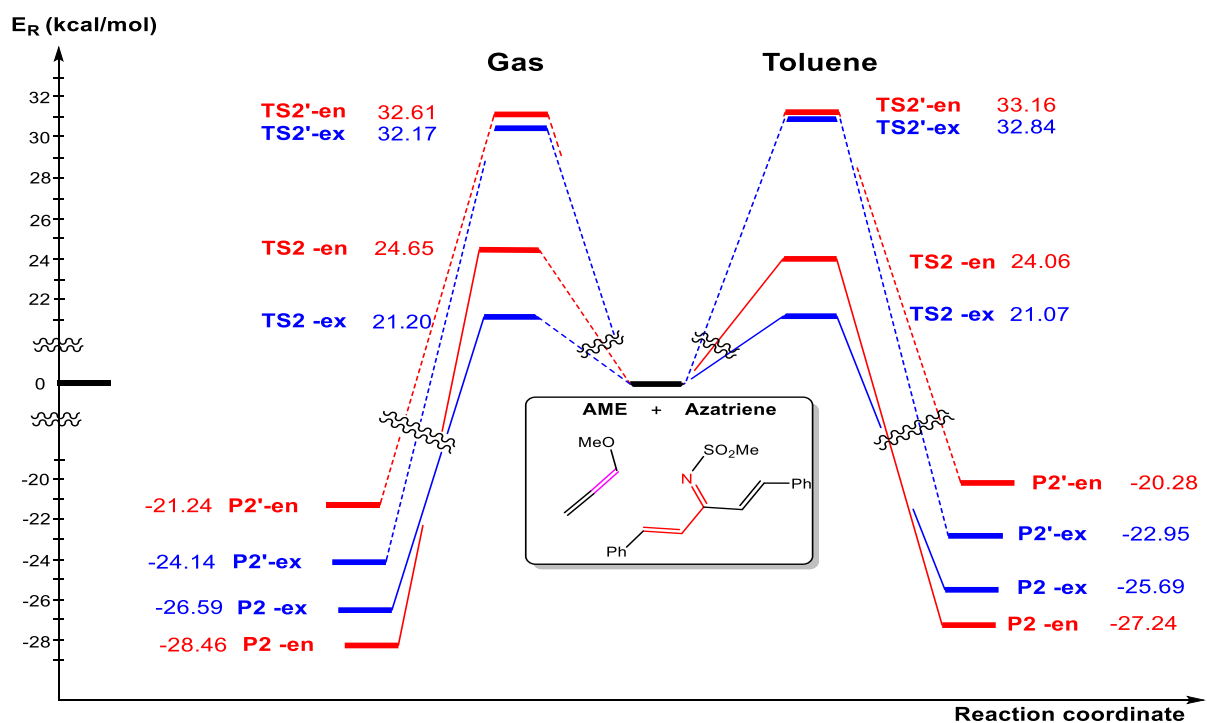

**Figure 4.** Activation energy profiles for the HDA reaction of **Azatriene** with **AME** in gas phase and in toluene.

Concerning the mechanism of the reaction, an IRC calculation was carried out for each TS, the located TSs were verified to correspond to the correct reaction path and to be unique through it. It is noteworthy that some of the TSs, in particular **TS2'-en** and **TS2'-ex**, have geometries (see **Figures 6** and **8**) that suggest the possibility of an intermediate in the reaction. The IRC plots corresponding to the **TS2'-en** and **TS2'-ex** have revealed what could be hidden intermediate regions[60]. Nevertheless, there was no success in locating an intermediate and, therefore, the possibility of a two-step mechanism can be disregarded. As we will discuss in detail in the next subsections, all reactions manifest a certain degree of asynchronicity in the new σ bonds formation process, which suggests that the reaction mechanism can still proceed according to more than one stage.

## 3. Mechanistic study
### 3.1. . Analysis of global electron density transfer

**Table 5** lists the GEDT values for the various TSs involved in the reaction of Azatriene with EVE and AME. Concerning the polarity of the reaction, we note that while most reaction pathways have GEDT values at the TS characteristic of polar reactions (i.e., with values higher than 0.20 e), the two transition states TS2'-en and TS2'-ex have GEDT values that are significantly shifted towards much lower values typical of low-polarity reactions. In this framework, the positive values of GEDT calculated at the level of the TSs for all the reactions indicate that the flow of electronic density takes place from nucleophilic sites of the dienophile to the electrophilic sites of dienes. Hence, the GEDT classifies the reactions as Reverse Electron Density Flux. We highlight that TS1-en and TS1-ex of the reaction of Azatriene with EVE present high GEDT values (0.35 e and 0.34 e respectively) associated with polar reactions and are typical of low activation barrier reaction pathways. This is consistent with the previous results summarized in the diagram of energy profiles (**Figure 4**). With regard to the reaction of Azatriene with AME we highlight that the highest GEDT value and consequently the highest polarity is observed for TS2-en and TS2-ex (0.30 e and 0.31 e). As a consequence, we expect that these reactions routes are be faster than the rest of pathways. A notable correlation between the polarity of a reaction and its rate has been documented before[61]. This is generally in harmony with the experimental results (assuming kinetic control of the reaction). Overall, the energetic order of TSs is consistent with the calculated GEDT and therefore with the polarity of reactions, i.e., the more pronounce the polar nature, the more accessible is the TSs.

**Table 5.** Global electron density transfer for the reaction of Azatriene with **EVE** and **AME** in the gas phase

| Reaction Azatriene + EVE | *TS1-en* | *TS1-ex* | *TS1'-en* | *TS1'-ex* |
|---|---|---|---|---|
| | 0,35 | 0,34 | 0,21 | 0,22 |
| Reaction Azatriene + AME | *TS2-en* | *TS2-ex* | *TS2'-en* | *TS2'-ex* |
| | 0.30 | 0.31 | 0.13 | 0.10 |

### 3.2. Analysis of Wiberg bond orders:

Wiberg Bond Order provides a way to calculate the percentage of bond formation and, thus, it allows to compare the degree of bond formation for the two forming σ bonds in a Diels-Alder reaction, as well as to determine the asynchronicity of bond formation and the mechanism of reaction. Concerning the reaction of **Azatriene** with **EVE**, we notice that for the two pathways **P1-en** and **P1-ex** relative to the first regioisomer the bond formations at the TSs are rather different. Indeed, the bond formation percentage of C1'-C1 bond (56.39%) advances that of C2'-N4 (21.38%) by about 35% (**Table 6**). These values show that these two pathways are characterized by an asynchronous bond formation process. On the contrary, for pathways **P1'-en** and **P1'-ex** relative to second regioisomer, we notice no significant difference in the percentage of bond formation of the two bonds C1'-N4 and C2'-C1 and hence the process is little asynchronous. Given these findings, and according to many studies linking high asynchronicity to lowering of reaction energy barriers, it can be deduced that for this reaction, the TSs corresponding to the more asynchronous processes i.e., pathways **TS1-en** and **TS1-ex**, are energetically more accessible TSs This result conforms to the previous results obtained by analysis of the different reaction energy pathways summarized in **Figure 2**. These results also support the proposition of an asynchronous one-step mechanism.

**Table 6.** Wiberg bond orders and percentage of the new bonds formation at the TSs of the reaction of Azatriene with EVE

| | **TS1-en** | **P1-en** | **TS1-ex** | **P1-ex** |
|---|---|---|---|---|
| *C2'-N4* | 0.199 | 0.929 | 0.182 | 0.933 |
| *(%)* | 21,38 | | 19,47 | |
| *C1'-C1* | 0.550 | 0.975 | 0.542 | 0.980 |
| *(%)* | 56,39 | | 55,31 | |
|  | **TS1'-en** | **P1'-en** | **TS1'-ex** | **P1'-ex** |
| *C1'-N4* | 0.395 | 0.947 | 0.419 | 0.943 |
| *(%)* | 41,73 | | 44,44 | |
| *C2'-C1* | 0.406 | 0.965 | 0.391 | 0.969 |
| *(%)* | 40,03 | | 40,33 | |

In the case of the reaction of **Azatriene** with **AME**, all four possible reaction pathways are revealed to be of asynchronous nature, all showing highly asynchronous bond formation. It is interesting to note that while the C1''-C1 bond is more advanced than C3''-N4 for regioisomer 1 (**TS2-en** and **TS2-ex**), this changes in the case of regioisomer 2 (**TS2'-en** and **TS2'-ex**) and it is the formation of the carbon-nitrogen C1''-N4 that is now more advanced (**Table 7**).

**Table 7.** Wiberg bond orders and percentage of the new bonds formation at the TSs of the reaction of Azatriene with **AME**

|  | **TS2-en** | **P2-en** | **TS2-ex** | **P2-ex** |
|---|---|---|---|---|
| *C3''-N4* | 0.175 | 0.937 | 0.149 | 0.905 |
| *(%)* | 18.64 | | 16.46 | |
| *C1''-C1* | 0.463 | 0.973 | 0.430 | 0.970 |
| *(%)* | 47.60 | | 44.36 | |
|  | **TS2'-en** | **P2'-en** | **TS2'-ex** | **P2'-ex** |
| *C1''-N4* | 0.488 | 0.977 | 0.480 | 0.976 |
| *(%)* | 49.98 | | 49.14 | |
| *C3''-C1* | 0.090 | 0.945 | 0.107 | 0.950 |
| *(%)* | 9.55 | | 11.28 | |

## 3.3. Topological analysis
### 3.3.1. AIM analysis of the inter-fragment interactions

The results of AIM analysis are reported in **Table 8**. For both reactions, examination of the electronic densities and Laplacians values at the TSs level reveals that the interatomic new bond regions experience significant non-covalent interactions. Indeed, all interatomic new bond regions manifest bond critical points characterized by a low magnitude electronic density and a positive Laplacian indicating a locally depleted density in the form of non-covalent interactions. These values of the electronic density also show harmony with previous conclusions on the asynchronous nature of bond formation observed for the TSs in the reactions of **Azatriene** with **EVE**, and of **Azatriene** with **AME**.

**Table 8.** Total electron density, ρ (a.u.) and Laplacian of electron density (a.u.) of the (3, -1) bond critical points at the TSs associated with the HDA reactions of Azatriene with **EVE** and **AME**

|  | $\rho$ | $\nabla^2\rho(r_c)$ | $\rho$ | $\nabla^2\rho(r_c)$ |
|---|---|---|---|---|
|  | CP (C1'-C1) | | CP (C2'-N4) | |
| **TS1-en** | 0.091 | 0.0004 | 0.033 | 0.0741 |
| **TS1-ex** | 0.090 | 0.0021 | 0.032 | 0.0706 |
| **TS1'-en** | 0.064 | 0.0339 | 0.067 | 0.0906 |
| **TS1'-ex** | 0.060 | 0.0403 | 0.072 | 0.0881 |
|  | CP (C1''-C1) | | CP (C2''-N4) | |
| **TS2-en** | 0.076 | 0.0238 | 0.030 | 0.0714 |
| **TS2-ex** | 0.085 | 0.0192 | 0.025 | 0.0625 |
| **TS2'-en** | 0.012 | 0.0356 | 0.106 | 0.0476 |
| **TS2'-ex** | 0.019 | 0.0496 | 0.104 | 0.0499 |

### 3.3.2. IGM analysis of the inter-fragment interactions

The IGMH results are analyzed in order to gain insights on the evolution of the electron density at the TS, as well as to evaluate the asynchronicity of the reactions. The results corresponding to the reaction of **Azatriene** with **EVE** and with **AME** are illustrated in **Figure 6** and **Figure 8** respectively. The IBSIW is also calculated as a complementary index to evaluate the interaction strength along the forming bonds at the TSs level. The relevant values of the IBSIW, together with the bond distances and the Wiberg indices of relevant bonds are shown in **Figures 7** and **9**.

For the reaction of **Azatriene** with **EVE**, it can be seen in **Figure 6** that the non-covalent interactions in the bonding region of all TSs are generally similar, however, **TS1-en** and **TS1-ex** exhibit an additional weak hydrogen bond between the oxygen of **EVE** and one of the hydrogens of the methyl group within the SO$_2$Me group of the **Azatriene**. This is indicated by a pale blue isosurface created in between the aforementioned oxygen and hydrogen atoms, and it more apparent in the case of **TS1-en** than for **TS1-ex**, which supports the regioselective and stereoselective nature of this reaction towards the endo product. The IBSIWs and Wiberg indices generally predict similar trends for the bond interaction strengths. For the reaction of **Azatriene** and **EVE**, a more pronounce asynchronous process is calculated for the first regioisomer (**TSI-en** and **TSI-ex**). In particular, the IBSIW value of C-C bond (17.1%) in **TS1-en** is significantly larger than the C-N bond (5.09%) (see Figure 7).

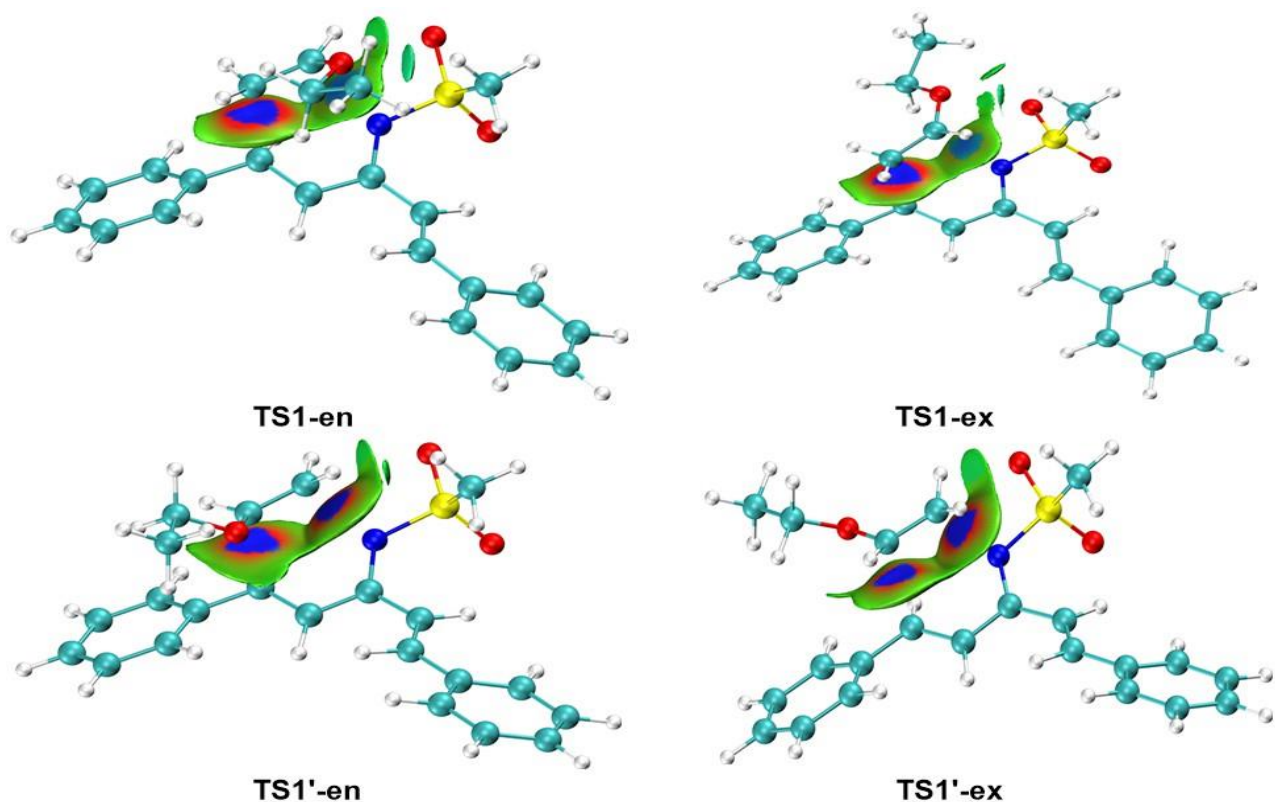

**Figure 5.** sign(λ2)ρ colored isosurfaces of δg$^{inter}$ = 0.008 a.u of TSs of reaction of **Azatriene** with **EVE**. Blue is for attractive interactions, red is for repulsive interactions, and green is for Van der Waals forces.

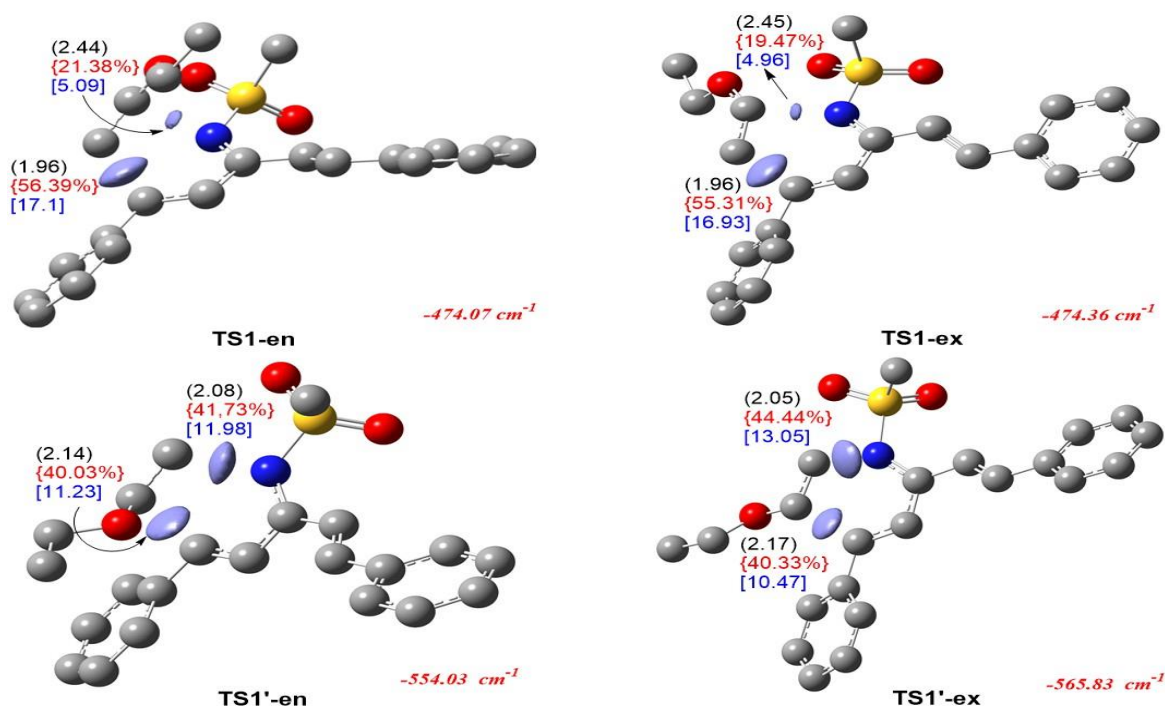

**Figure 6.** New bond distances in Å (in parentheses), percentage of the bond formation based on Wiberg indices (in curly brackets), and IBSIWs (in square brackets) for TSs of the reaction of **Azatriene** with **EVE.**

A similar trend is observed in Figure 8 for **TS2-en** and **TS2-ex** in the reaction of **Azatriene** with **AME**, where we again observe weak though favorable interactions that stabilize the TSs. Furthermore, as previously mentioned, **TS2′-en** and **TS2′-ex** exhibit a highly asynchronous nature characterized by geometries that are similar to what we would expect in an intermediate. Indeed, while one bond (C2''-N4) has already begun forming, with a corresponding bond length of 1.86 Å and 1.88 Å for **TS2'-en** and **TS2'-ex** respectively, the second one (C3''-C1) has not, and the corresponding bond length is 3.09 Å and 2.78 Å. This feature emphasizes the two-stage nature of the reaction mechanism[30]. The origin of the features characterizing these TSs can be explored through analysis of the electronic density associated with weak interactions (see Figure 8). In the case of **TS2'-ex**, a localized isosurface of green shade is observed in the space between the oxygen of the SO$_2$Me group and one of the hydrogens of the OMe group of **AME**, and which can be associated with a weak hydrogen bond. This weak interaction is characterized by an IBSIW index of 3.87 (see Figure 9), and appears to result in the orientation of the C3''OMe terminal backbone of **AME** in such a way that it becomes difficult for the C3'' carbon to assume a bonding orientation with the C1 carbon at this stage of the reaction. A similar though less intense interaction is observed for **TS2'-en**, in this case the weak interaction is established between the SO$_2$Me group and the hydrogen on the C3'' of **AME**, it is characterized by an IBSIW of 0.83 (not visible in **Figure 9**) and associated with a Van der Waals interaction. It responsible, at least in part, for the geometry adopted by the TS. The high activation barriers associated with **TS2'-en** and **TS2'-ex** may be attributed to their distorted geometries, therefore, their relative instability partially arises from weak interactions that disfavor the anticipated bonding patterns. Both **TS2-en** and **TS2-ex** show significant attractive non-covalent interactions in the bonding region in comparison with the other TSs, as observed by the large blue isosurfaces in Figure 8. Again, this reflects the competing nature between the endo and the exo products of the reaction of **Azatriene** and **AME**. Nevertheless, in this later situation, it is difficult to unveil which of the two TSs is the more stable.

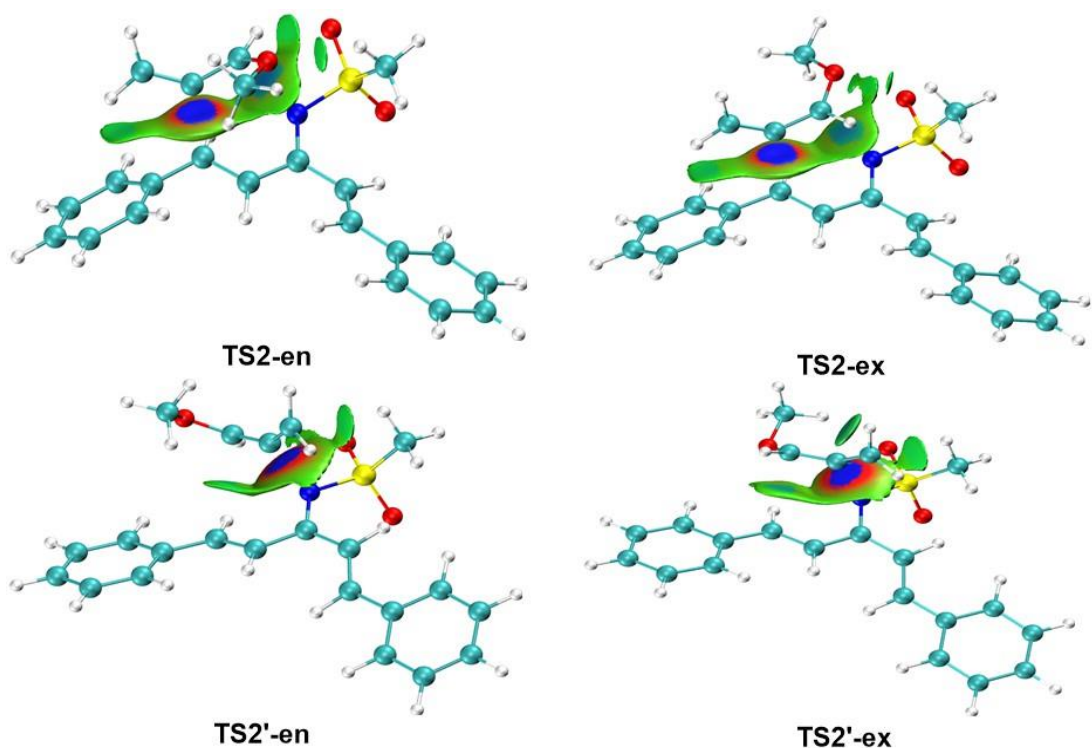

**Figure 8.** sign($\lambda2$)ρ colored isosurfaces of δg$^{inter}$ = 0.008 a.u of TSs of reaction of **Azatriene** with **AME**. Blue is for attractive interactions, red is for repulsive interactions, and green is for Van der Waals forces.

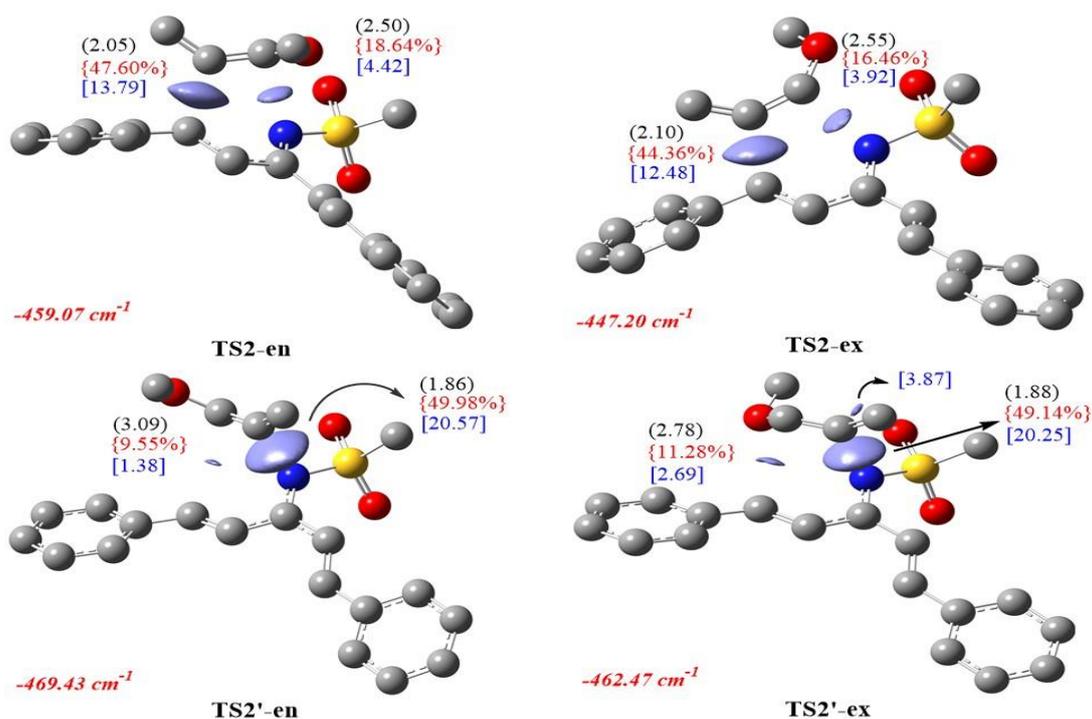

**Figure 9:** New bond distances in Å (in parentheses), percentage of the bond formation based on Wiberg indices (in curly brackets), and IBSIWs (in square brackets) for TSs of the reaction of **Azatriene** with **AME.**

### 3.3.3. ELF topological analysis at the ground state of TSs

The electronic density of the TSs of the reactions of **Azatriene** with **EVE** and **AME** is investigated from the perspective of ELF topological analysis. The results are presented in **Figure 10** and **Figure 11** for the reaction with **EVE** and **AME**, respectively. All TSs show the expected trend where the double bonds involved in the DA reaction, i.e., bonds C1'-C2', C1-C2, and C3-N4, are depopulated in favor of the population of the forming σ bonds (C1'-C1/C2 and C2'-C2/C1) as well as the forming π bond in the diene fragment. This behavior is more pronounced in the case of **TS1-en** and **TS1-ex**, and goes accompanied by the appearance of a disynaptic basin V(C1',C1) integrating a population of 0.89 e. The latter indicates that **TS1-en** and **TS1-ex** are located at a stage of the reaction path where the new bond C1'-C1 is already forming. In the case of TSs **TS1'-en/ex,** unlike with **TS1-en/ex,** we observe one monosynaptic basin V(C2') integrating 0.42 e for **TS1'-en** and two monosynaptic basins V(C2') and V(C1') integrating 0.37 e and 0.06 e for **TS1'-ex**. The monosynaptic basins V(C2') are associated with pseudoradical centers. This signifies that the TSs are at a stage where the new σ bonds formation have not practically started. From an asynchronicity perspective, an important electron density is accumulated in the C-C bond region for all TSs, especially **TS1-en** and **TS1-ex**, whereas the C-N bond region does not show any noteworthy density. This endorses the asynchronous character of the reaction mechanism. It can also be observed that all emerging basins in the bonding region are related to the dienophile, thus the electron flux is directed from the dienophile to the diene. It is interesting to note that the emerging basins at the bonding regions are located between the most Nucleophilic center of the **EVE** and the most electrophilic center of the **Azatriene** for **TS1-en** and **TS1-ex**. Nevertheless, for **TS1'-en** and **TS1'-ex** the emerging basins can only be associated with the most electrophilic center located at the diene.

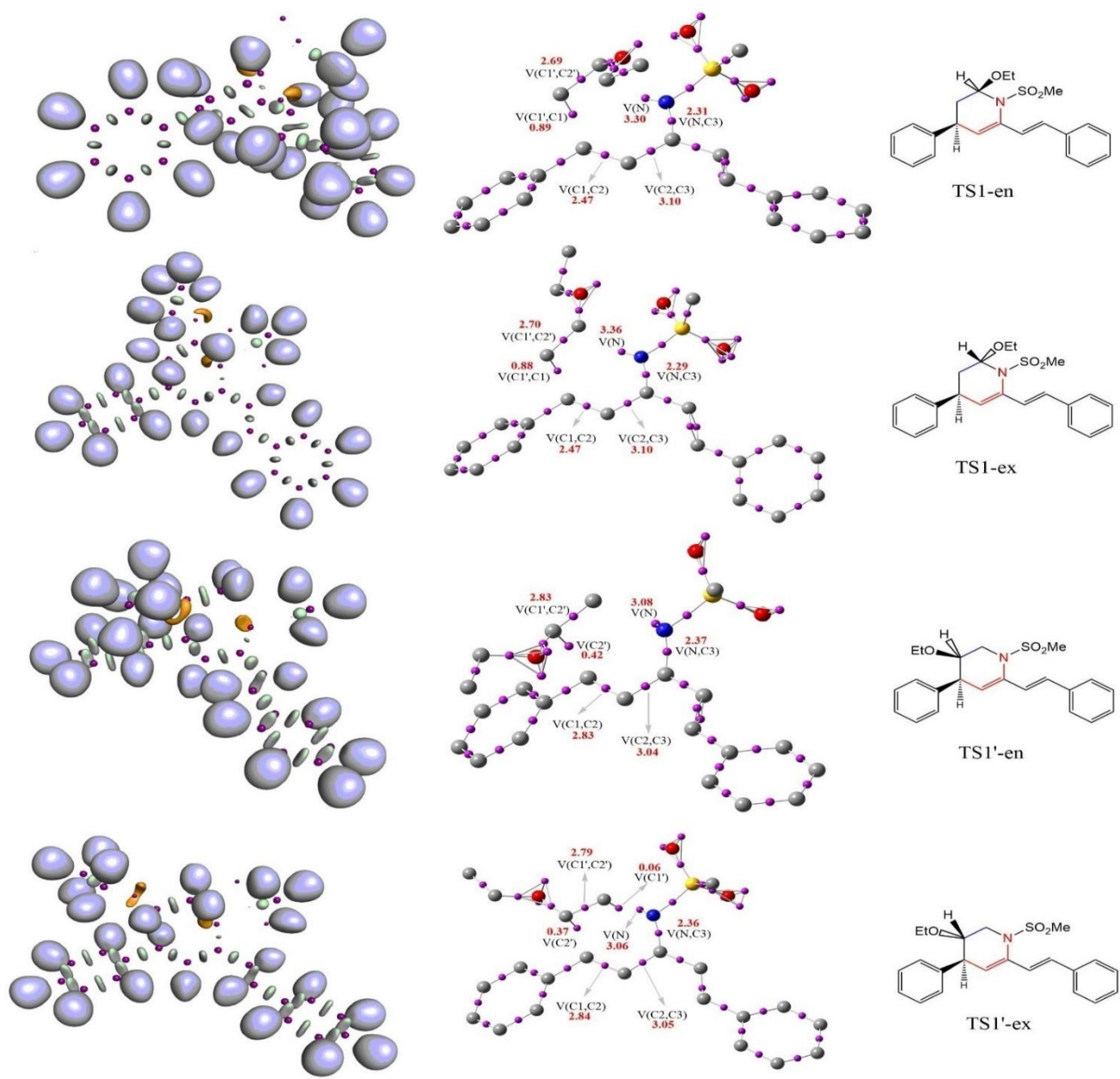

**Figure 10.** (Left) ELF localization domains of the TSs associated with the reaction of **Azatriene** and **EVE** plotted at an isosurface value of 0.8. Protonated basins are shown in light blue, monosynaptic basins in orange, disynaptic basins in green and core basins in purple. (Middle) ELF basin attractor positions along with population values for the most relevant sites. (Right) Lewis-like structures of reactants.

Similar to the reaction of **Azatriene** with **EVE**, the electronic density in the reaction of **Azatriene** with **AME** evolves in a way that depopulates the double bonds involved in the DA reaction in favor of the population of the forming bonds.

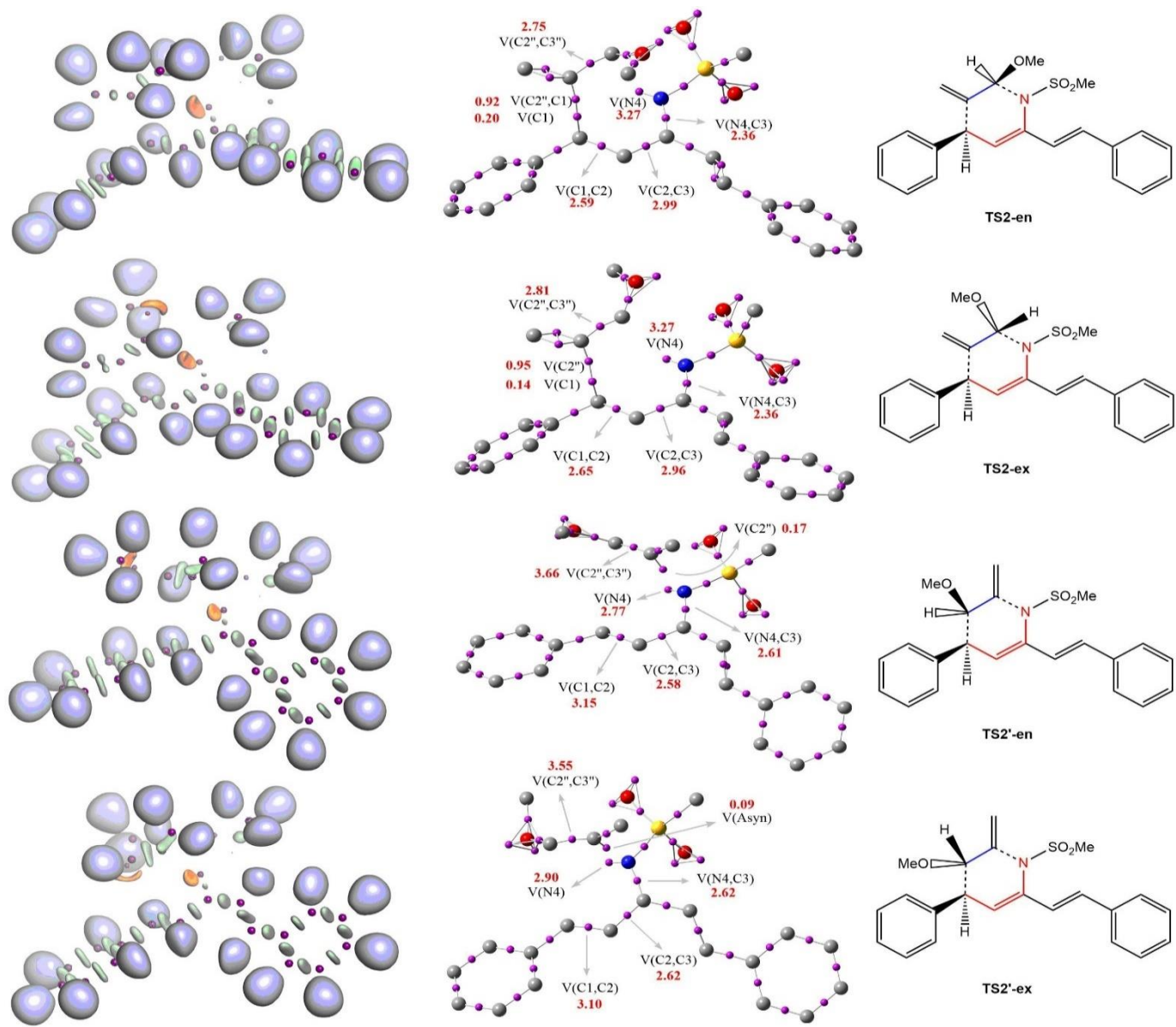

**Figure 11.** (Left) ELF localization domains of the TSs associated with the reaction of **Azatriene** and **AME** plotted at an isosurface value of 0.8. Protonated basins are shown in light blue, monosynaptic basins in orange, disynaptic basins in green and core basins in purple. (Middle) ELF basin attractor positions along with population values for the most relevant sites. (Right) Lewis-like structures of reactants.

For all TSs, the most remarkable changes occur in the dienophile **AME** where the population of the emerging basins is considerably higher than those emerging from the diene's side. In fact, **TS2'-en** and **TS2'-ex** in Figure 11 do not exhibit any basins emerging from the diene's side, this attests to the REDF nature of the reaction. In addition, it can be clearly seen that the electronic density of **TS2-en** and **TS2-ex** is different from that of **TS2'-en** and **TS2'-ex**. In **TS2-en** and **TS2-ex**, similar to the reaction of **Azatriene** with **EVE**, the C-C bond is in the process of forming, while the C-N lags behind. On the contrary, in **TS2'-en** and **TS2'-ex**, the C-N bond formation is ahead of that of the C-C bond. Examination of the population of the basins in the C-N bond region shows a monosynaptic basin V(C1) integrating a very small population of 0.17 e and 0.09 e for **TS2'-en** and **TS2'-ex,** respectively., this means that these transition structures are located at a stage where the bond formation has not started yet. From the perspective of nucleophilicity and electrophilicity, we observe that for **TS2-en** and **TS2-ex**, the forming density is associated simultaneously with the most electrophilic and nucleophilic centers of the diene and the dienophile respectively, although the same cannot be said about **TS2'-en** and **TS2'-ex**. This observation, together with the similar observation for the reaction of **Azatriene** with **EVE** leads us to conclude that transition structures with low activation barriers (**TS1-en/ex** and **TS2-en/ex**) tend to involve the population of bonds that include simultaneously the most electrophilic and the most nucleophilic centers of the two reactants. In contrast, the transition structures with high activation barriers (**TS1'-en/ex** and **TS2'-en/ex**) do not satisfy the previous requirement.

From the data obtained in the different approaches, it becomes evident that all the reaction pathways studied in this work are asynchronous mechanisms where the new bonds formation occurs in two distinct stages, i.e., in a one-step two-stages mechanism. This asynchronicity though is not such that that the first stage involves a separable intermediate, and the mechanism is still overall considered a one-step mechanism.

## Conclusion

The reactions of diene N-sulfonyldivinylmethanimine with the dienophiles ethyl vinyl ether and allenyl methyl ether are investigated. A good correlation between the electronic structure calculations and the experimentally demonstrated reactivity is established. The energies of the reactions, especially the activation barriers are coherent with the findings of the topological, Wiberg bond order, and GEDT analyses, and show a clear relationship with the characteristics of the reactions mechanism. We find that the reactions are carried out under kinetic control. Overall, the reactions are polar though a few reaction pathways in the reaction of N-sulfonyl divinyl methanimine with allenyl methyl ether manifest a low polar character. The electronic flux takes place from the dienophiles to the diene. The mechanism is found to be an asynchronous one-step two-stage mechanism, and despite suspicion over the involvement of a stepwise mechanism in in the particular case of the reaction of N-sulfonyldivinylmethanimine with allenyl methyl ether, all attempts to locate intermediates were futile. CDFT indices correctly capture the effect of the substituents in N-sulfonyldivinylmethanimine on its reactivity. Indeed, they permit classifying the various substituted dienes according to their increasing reactivity with the dienophiles of this work, and it is shown that their reactivity is depends on their ability to reinforce the reverse electron density flux of the reaction. Furthermore, CDFT indices successfully predict the regiochemistry and periselectivity of the reactions. The endo product of the reaction of N-sulfonyldivinylmethanimine with ethyl vinyl ether is the preferential one and is favored by the establishment of a weak hydrogen bond between the oxygen of ethyl vinyl ether and one of the hydrogens of the methyl group within the $SO_2Me$ group. The exo approach of the reaction of N-sulfonyldivinylmethanimine with allenyl methyl ether is the preferential. The calculated energy difference between the endo and exo products in the case of the reaction of N-sulfonyldivinylmethanimine and ethyl vinyl ether is not enough to explain the experimental ratios of the endo/exo products. We expect that a further improvement in quantum chemical methods will help us achieve a full predictive potential. The results obtained in this work shed light on the available experimental results and can be considered as a solid ground for further research on the reactivity of azatrienes in hetero-Diels-Alder reactions as well as their products.

## Acknowledgements

This work has received funding from the European Union's Horizon 2020 research and innovation program under Marie Sklodowska-Curie grant agreement No. 872081 and grant PID2022-136228NB-C21 funded by MICIU/AEI/10.13039/501100011033 and, as appropriate, by "ERDF A way of making Europe", by "ERDF/EU", by the "European Union" or by the "European Union NextGenerationEU/PRTR". The authors acknowledge the Moroccan Association of Theoretical Chemists for providing the computational programs.

## Author Contribution Statement

The authors confirm contribution to the paper as follows:
**Amine Rafik**: Conceptualization, Investigation, Data curation, Methodology, Formal analysis, Visualization, Software, original draft, Review & editing. **Abdeljabbar Jaddi**: Methodology, Formal analysis, Visualization. **Najia Komiha**: Conceptualization, Formal analysis, Investigation. **Mohammed Salah**: Formal analysis, Investigation, Validation. **Miguel Carvajal**: Formal analysis, Review & editing, Supervision, Validation, Funding acquisition. **Khadija Marakchi**: Conceptualization, Investigation, Methodology, Formal analysis, Review & editing, Supervision, Validation, Software.